\documentclass[structabstract]{aa}  
\usepackage{natbib}
\bibpunct{(}{)}{;}{a}{}{,} 
\usepackage{graphicx}

\usepackage{txfonts}
\begin{document}

\title{Environmental Dependence of Local Luminous Infrared Galaxies}
\author{Ho Seong Hwang \inst{1}
\and David Elbaz \inst{1}
\and Jong Chul Lee \inst{2}
\and Woong-Seob Jeong \inst{3}
\and Changbom Park \inst{4}
\and Myung Gyoon Lee \inst{2}
\and Hyung Mok Lee \inst{2}
}
\institute{CEA Saclay/Service d'Astrophysique, F-91191 Gif-sur-Yvette, France \\
\email{hoseong.hwang@cea.fr}
\and Astronomy Program, Department of Physics and Astronomy, Seoul National University,
   Seoul 151-742, Korea 
\and Space Science Division, Korea Astronomy \& Space Science Institute, 
  Deajeon 305-348, Korea 
\and School of Physics, Korea Institute for Advanced Study, Seoul 130-722, Korea 
}

\date{Received September 15, 1996; accepted March 16, 1997}

 
\abstract
{}
{We study the environmental dependence of local luminous infrared galaxies (LIRGs)
  and ultraluminous infrared galaxies (ULIRGs)
  found in the Sloan Digital Sky Survey (SDSS) data.}
{The LIRG and ULIRG samples are constructed by cross-correlating
  spectroscopic catalogs of galaxies of the SDSS Data Release 7
  and the Infrared Astronomical Satellite Faint Source Catalog.
We examine the effects of the large-scale background density ($\Sigma_5$),
  galaxy clusters, and 
  the nearest neighbor galaxy on the properties of infrared galaxies (IRGs).
}
{We find that the fraction of LIRGs plus ULIRGs among IRGs ($f_{\rm (U)LIRGs}$)
  and the infrared luminosity ($L_{\rm IR}$) of IRGs
  strongly depend on the morphology of and the distance to the nearest neighbor galaxy:
  the probability for an IRG to be a (U)LIRG ($f_{\rm (U)LIRGs}$) and its $L_{\rm IR}$
  both increase as it approaches a late-type galaxy, but
  decrease as it approaches an early-type galaxy 
  (within half the virial radius of its neighbor).
We find no dependence of $f_{\rm (U)LIRGs}$ on 
  the background density (surface galaxy number density)
  at fixed stellar mass of galaxies.
The dependence of $f_{\rm (U)LIRGs}$ on the distance to galaxy clusters
  is also found to be very weak, but in highest-density regions such as 
  the center of galaxy clusters, few (U)LIRGs are found.
}
{These environmental dependence of LIRGs and ULIRGs and 
  the evolution of star formation rate (SFR)-environment relation 
  from high redshifts to low redshifts 
  seem to support the idea that galaxy-galaxy interactions/merging 
  play a critical role in triggering the star formation activity of 
  LIRGs and ULIRGs.}
{}


\authorrunning{H. S. Hwang et al.}
\maketitle
%

\section{Introduction}

Since the launch of the {\it Infrared Astronomical Satellite} 
  \citep[{\it IRAS}]{neu84} in 1983,
  a great deal of effort has been made to understand the nature
  of luminous infrared galaxies (LIRGs; $10^{11}\leq L_{\rm IR}<10^{12} L_\odot$)
  and ultraluminous infrared galaxies (ULIRGs; $L_{\rm IR}\geq 10^{12} L_\odot$):
(1) what triggering mechanism is responsible for the enormous infrared (IR) luminosity 
  (internally or externally driven?);
(2) what determines the relative importance between star formation (SF) and 
  active galactic nuclei (AGN) activity within (U)LIRGs
  (see \citealt{sm96,gc00,lon06,soi08} and references therein).
For example, most ULIRGs are found to be interacting systems 
  between two or more late-type galaxies (e.g., \citealt{dckim02,vei02}).
Many LIRGs are also found associated to interacting systems,
  but a significant fraction of them do not show definite features 
  of major mergers
  (\citealt{ish04,wang06}; see also \citealt{shi06,shi09,elb07,lotz08,ide09}).
Thus, minor mergers and galaxy-galaxy interactions may also play
  an important role in triggering the star formation activity (SFA) of LIRGs.

As the Sloan Digital Sky Survey (SDSS; \citealt{york00})
  and Two Degree Field Galaxy Redshift Survey (2dFGRS; \citealt{col01})
  have produced unprecedentedly large photometric and spectroscopic data 
  of nearby galaxies,
  the important role of the environment in determining the SFA 
  of galaxies in the local universe has been revealed and extensively studied
  (e.g., \citealt{lew02,gom03,kau04,bal04,tan04,owe07,park07,li08,pc09,ph09,hl09,bm09,jhlee10env}).
However, there are few studies focusing on how (U)LIRGs,
  showing much larger SFA compared to normal galaxies, 
  are affected by the environment.

\citet{goto05ir} investigated the optical properties of 
  IR galaxies 
  in the SDSS Data Release 3 (DR3) and found that 
  more luminous IR galaxies tend to be located in lower density regions.
Later, \citet{zau07} studied the environment of local ULIRGs at $z<0.3$
  using the spatial correlation amplitude parameter $B_{\rm gc}$.
The environment of most ULIRGs was found to be similar to that of the field,
  but a small number of ULIRGs was found in high-density regions such as galaxy clusters.
They also found no significant difference in the $B_{\rm gc}$ distribution of local ULIRGs
  with those of local Seyferts, local QSOs, and intermediate-z QSOs,
Recently, \citet{wr10lf}, 
  using the Imperial IRAS-FSC redshift catalog \citep{wr09cat},
  investigated the spatial clustering of galaxies as a function of
  several galaxy properties.
They found that cirrus-type galaxies (dust emission due to 
  the interstellar radiation field) show a stronger clustering than 
  M82-like starbursts (interaction-induced starbursts),
  which indicates that the latter preferentially resides 
  in low-density environments.
They also found a correlation between the correlation strength and the IR
  luminosity for M82-like actively star-forming galaxies 
  except in high-density regions such as the center of galaxy clusters.

The environment of (U)LIRGs beyond the local universe has recently
  started to be explored.
\citet{far06} studied a spatial clustering of ULIRGs at $1.5<z<3$, and found that 
  these ULIRGs cluster more strongly than other galaxies at these epochs.
Similarly, \citet{gil07} suggested that 
  $24\mu$m-selected, star-forming galaxies at $z\sim1$ 
  in the Great Observatories Origins Deep Survey (GOODS) appear to be
  more clustered than optically selected galaxies at similar redshifts,
  and that more luminous IR galaxies have a larger correlation length.
Recently, \citet{mar08} investigated the environment of LIRGs and ULIRGs at $0.7<z<1$
  in the Extended Groth Strip (EGS).
They found that the local environment of (U)LIRGs is intermediate
  between those of red and blue galaxies.
If they compare (U)LIRGs with non-IR galaxies whose stellar mass,
  color, or luminosity are similar to those of the (U)LIRGs,
  they see no significant difference in the environment between the two.

Since LIRGs and ULIRGs dominate the SFA at high redshifts (e.g., \citealt{floch05,bmag09}),
  the understanding of the nature of (U)LIRGs is closely related to
  the study of the evolution of the star formation rate (SFR)-environment relation.
For example, \citet{elb07}, using the GOODS data, 
  discovered that the SFR-density relation observed locally
  is reversed at $z\sim1$, in the sense that the ``average'' SFR of galaxies over 
  a given volume increases with the local galaxy density 
  (see also \citealt{coo08,ide09,tran10}).
On the contrary, \citet{pat09} observed the galaxies in a galaxy cluster 
  (RX J0152.7$-$1357) at $z=0.834$,
  and found that the SFR-density relation at that redshift is not
  similar to those in \citet{elb07} and \citet{coo08}, but similar to
  that in the local universe (see also \citealt{fer09}).
It is noted that the SFR in their study is the ``median'' SFR in a given mass bin,
  which is different from the spatially ``averaged'' SFR of the nearby galaxies
  in \citet{elb07} and \citet{coo08}.
\citet{pat09} argued that the contradictory results may be caused by
  the different environment that they investigate and 
  by the different sample selection.
They observed the central region of galaxy clusters
  (that is not well resolved by the other studies).
Moreover, they used a mass-limited sample of galaxies
  that is different from the luminosity-limited sample of galaxies used
  in other studies, and the latter can be affected by the contribution of 
  low-mass, blue star-forming galaxies.

In this paper, 
  we study the environmental dependence of (U)LIRGs found in the recent SDSS data
  to improve understanding of the triggering mechanism of (U)LIRGs' SFA
  and of  the evolution of the SFR-environment relation.
Section \ref{data} describes the data and the environmental indicators used in this study.
Environmental dependence of the physical properties is given in \S \ref{results}.
Discussions and conclusions are given in \S \ref{discuss} and \S \ref{con}, respectively.
Throughout, we adopt $h=0.7$ and a flat $\Lambda$CDM cosmology with density parameters 
  $\Omega_{\Lambda}=0.73$ and $\Omega_{m}=0.27$.
  

\section{Data}\label{data}
\subsection{Observational Data Set}

To identify LIRGs, we first cross-correlate the largest available 
  IR source catalog and redshift catalog.
For the IR source catalog,
  we use the {\it IRAS} Faint Sources Catalog -- Version 2 (\citealt{mos92}, 
  hereafter FSC92), which contains 173,044 IR sources at $\mid b \mid > 10 \deg$
  having measured fluxes at 12, 25, 60 and 100 $\mu$m.

For the redshift catalog, we use a spectroscopic sample of galaxies
  including the main galaxy sample ($m_r<17.77$) and 
  faint galaxies ($m_r>17.77$) whose spectroscopic redshifts are available 
  in the SDSS DR7 \citep{aba09}.
Completeness of the spectroscopic data in SDSS is poor
  for bright galaxies with $m_r<14.5$ because of
  the problems of saturation and cross-talk in the spectrograph, and
  for the galaxies located in high-density regions such as galaxy clusters
  due to the fiber collision.
Thus, it is necessary to supplement the missing galaxy data
  to reduce the possible effects of the incompleteness problem.
Therefore, we also use a photometric sample of galaxies with $m_r<17.77$
  whose redshift information is available in the literature.
We download the photometric sample of galaxies with $m_r<17.77$
  and match it with the redshift catalogs of galaxies
  such as Updated Zwicky Catalog (UZC; \citealt{fal99}),
  MX northern Abell cluster redshift survey \citep{sli98},
  EFAR survey \citep{weg96,weg99},
  the redshift survey of Abell clusters \citep{ho93,ho98}.
Furthermore, we add redshift information of the photometric sample of galaxies
  located within ten times the virial radius of some galaxy clusters 
  from the NASA Extragalactic Database (NED) \citep{ph09}.
In total, the redshift information for 184,845 galaxies in the photometric sample
  is compiled, which is overlapped with 173,828 galaxies in the spectroscopic sample. 
Finally, we add 11,017 galaxies to the spectroscopic sample of galaxies,
  which yields a final sample of 929,234 galaxies.
In the result, the spectroscopic completeness of our sample
  is higher than $85\%$ at all magnitudes with $m_r<17.77$ 
  and even in the center of galaxy clusters (see Fig. 1 in \citealt{ph09}).
It is to be noted that the final sample also contains the galaxies 
  whose spectral classifications provided in the SDSS, are QSOs.

In order to investigate the physical parameters of galaxies,
  we use several value-added galaxy catalogs (VAGCs) drawn from SDSS data.
Photometric and structure parameters are adopted from SDSS pipeline \citep{sto02}.
The velocity dispersion of galaxies is adopted from
  DR7 release of VAGC \citep{bla05}.
The stellar mass estimates are obtained from 
  MPA/JHU DR7 VAGC\footnote{http://www.mpa-garching.mpg.de/SDSS/DR7/},
  which is based on the fit of SDSS five-band photometry 
  with the model of \citet{bc03} \citep{kau03,gal05}.
The $H\alpha$ fluxes are also taken from MPA/JHU DR7 VAGC,
 which is computed from Gaussian fit to continuum subtracted data \citep{tre04}.
  
We adopt the galaxy morphology information from
  Korea Institute for Advanced Study (KIAS) DR7 VAGC 
  (\citealt{pc05,choi07}; Choi 2010 in prep.),
  which contains 697,320 main galaxy sample in NYU VAGCs as well as
  10,497 photometric sample of galaxies with the redshift information 
  from various existing redshift catalogs.
We perform additional visual classification 
  for the galaxies in DR7 that are not included in KIAS DR7 VAGC.
During the visual inspection of color images of galaxies,
  we eliminate 403 spurious sources (e.g., faint fragments of bright galaxies, 
  diffraction spikes of bright stars), and
  they are not included in the final sample of 929,234 galaxies.

\subsection{The Cross-Correlation}

The positional uncertainties of the {\it IRAS\/} sources are
  different depending on the scan direction
  (typically 5$''$  in-scan direction and 16$''$ for the
  cross-scan direction, 1 $\sigma$), and vary from source to source 
  (1$''-$13$''$ for the in-scan direction, or minor axis of
  positional uncertainty ellipse and 3$''-$55$''$ for the cross-scan
  direction, or major axis of positional uncertainty ellipse, 1 $\sigma$). 
The positional uncertainties of the optical counterparts
  are negligible in comparison. 
If a galaxy in SDSS lies within 3 $\sigma$ positional uncertainty ellipse 
  of the {\it IRAS} source, we regard it a match.
It gave us 15,611 {\it IRAS} sources having optical counterparts in the SDSS.
When there are more than one SDSS galaxy within the {\it IRAS} ellipse
 (10\% of the total number of matchings),
  we chose the one closest to the center of the {\it IRAS} source.

\begin{figure}
\center
\includegraphics[scale=0.42]{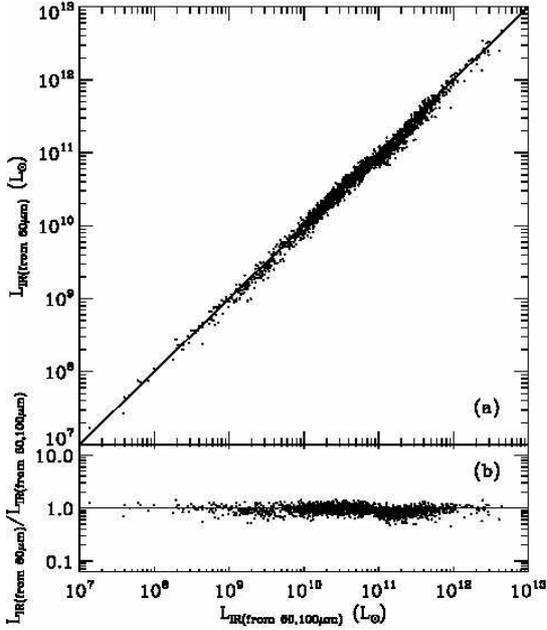}
\caption{Comparison of IR luminosities ($L_{\rm IR}$) 
  computed from the {\it IRAS} 60 $\mu$m fluxes only ($L_{\rm IR(from~60\mu m)}$) with 
  those from the {\it IRAS} 60 and 100 $\mu$m ($L_{\rm IR(from~60,100\mu m)}$).
We only show 30\% of the total sample.
}\label{fig-lum}
\end{figure}

We use the spectral energy distribution (SED) models of \citet{ce01} to compute 
  the IR luminosity of galaxies.
We restrict the computation to 15,547 sources whose {\it IRAS} 60 $\mu$m fluxes are reliable,
  which means that the flux quality flags are 
  either `high quality' or `moderate quality'.
Fluxes at 60 and 100 $\mu$m
  are used for the SED fit when their fluxes are reliable.
If 100 $\mu$m flux is not reliable, we use only 60 $\mu$m flux for the SED fit.
It was pointed out that 100 $\mu$m flux of some IRAS sources 
  could be overestimated \citep{jeong07iras}, but
  its effect on the estimation of IR luminosity is expected to be less than a factor of two. 
Since flux quality flags are different depending on the sources, we show
   how the derived IR luminosities agree with each other in Fig. \ref{fig-lum}.
It shows that IR luminosities computed from 
  the {\it IRAS} 60 and 100 $\mu$m ($L_{\rm IR(from~60,100\mu m)}$)
  and those from the {\it IRAS} 60 $\mu$m fluxes only
  ($L_{\rm IR(from~60\mu m)}$) agree well with rms=0.08 in a logarithmic scale.
We further reject 40 galaxies whose redshift confidence parameters (zConf)
  are smaller than 0.65, and finally have a sample of 15,507 IR galaxies.

\begin{figure}
\center
\includegraphics[scale=0.45]{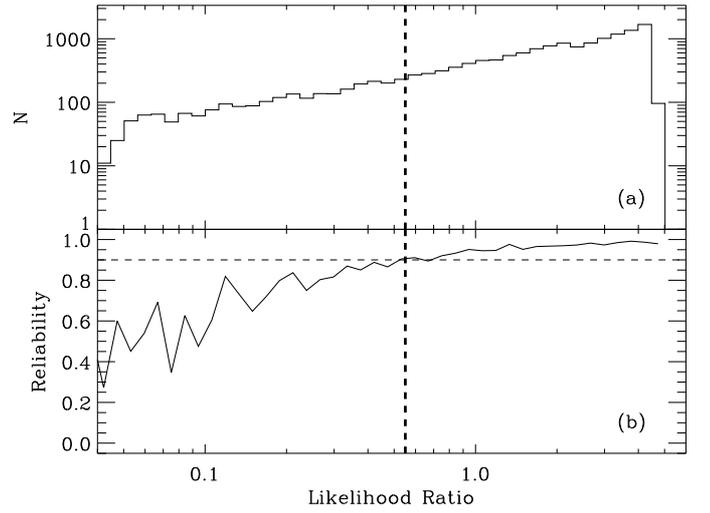}
\caption{(a) Distribution of likelihood ratio (LR). 
(b) Reliability of our cross-correlation vs. LR. 
Vertical dashed line represents the critical LR values (90\%) of 
  reliable identification for {\it IRAS} sources adopted in this study.
}\label{fig-lr}
\end{figure}

Since our sample is constructed by the cross-correlation based on
  the position alone, it may contain spurious sources due to the chance
  presence of a galaxy within the positional uncertainty of IR sources.
In order to secure the sample of highly reliable association
  between IR sources and the optical counterparts, 
  we compute the likelihood ratio (LR) for each association \citep{ss92}.
The LR is defined by the ratio of the probability 
  of a true association to that of a chance association\footnote{We assume 
  that the positional errors of {\it IRAS} sources are Gaussian.}
\begin{equation}
{\rm LR} = \frac{Q(\leq m) \exp(-r^{2}/2)}{2\pi\sigma_{\rm 1}\sigma_{\rm 2}n(\leq m)}, \label{eq-LR}
\end{equation}

where $n(\leq m)$ is the local surface density of objects brighter than the candidate. 
The ``normalized distance" $r$ is given by
\begin{equation}
r^{2} = \left( \frac{d_1}{\sigma_{1}} \right)^{2} 
  +\left(\frac{d_2}{\sigma_{2}} \right)^{2}, \label{eq-normdist}
\end{equation}
where $d_1$ and $d_2$ are positional differences along the
  two axes of an error ellipse for an IR source, and
  $\sigma_1$, $\sigma_2$ are the lengths of these axes. 
Since the positional uncertainties of SDSS galaxies are negligible 
  compared to those of {\it IRAS} sources, 
  we define $\sigma$ as the length of the error axes of the {\it IRAS} sources. 
$Q(\leq m)$ is a multiplicative factor which is a prior
  probability that a ``true'' optical counterpart brighter than the
  magnitude limit exists in the association, and for simplicity we
  set $Q=1$ in this study.

\begin{figure}
\center
\includegraphics[scale=0.45]{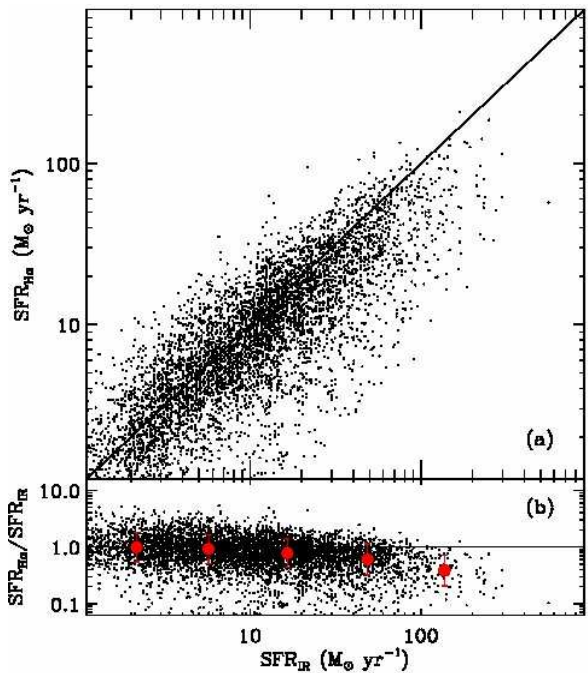}
\caption{Comparison of SFRs determined from $L_{\rm IR}$ with those from $L_{\rm H\alpha}$.
We only plot galaxies whose spectral type is classified as star-forming (see \S \ref{parameter}).
Filled circles in (b) are median values of the ratios (SFR$_{\rm H\alpha}$/SFR$_{\rm IR}$)
  at each bin, and the error bars indicate the dispersion of the ratios.
}\label{fig-sfr}
\end{figure}

\begin{figure}
\center
\includegraphics[scale=0.45]{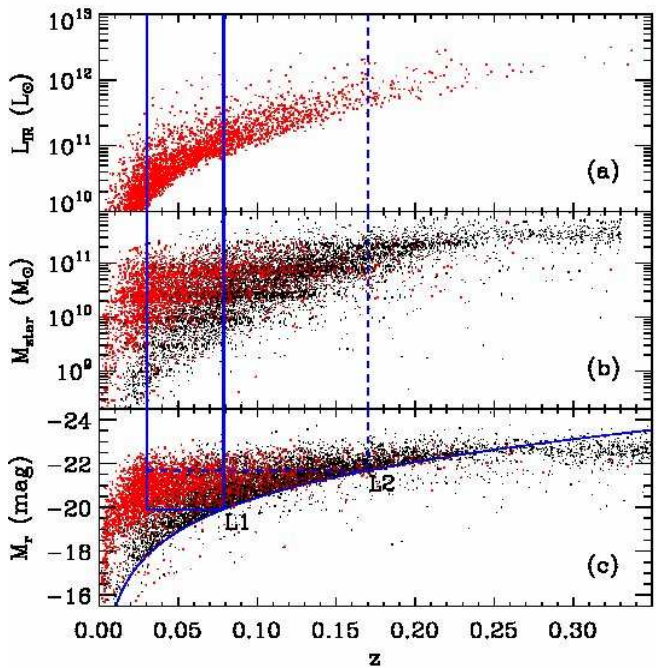}
\caption{(a) IR luminosity, (b) stellar mass, and (c) absolute $r$-band magnitude vs. redshift. 
Red dots indicate IRGs (only 30$\%$ of IRGs are shown), and black dots denote
  galaxies without IR detection in the spectroscopic sample
  (only 1 $\%$ of galaxies in the total sample are shown).
Solid and dashed lines in (c) define the volume limited samples, 
  L1 and L2, respectively.
The bottom curve corresponds to the apparent magnitude limit ($m_r=17.77$)
  for the main galaxy sample in SDSS.
}\label{fig-vol}
\end{figure}

We compute $n(\leq m)$ using the photometric sample of galaxies 
  within 3 $\sigma$ error ellipses,
\begin{equation}
n(\leq m) = \frac{N(\leq m)}{\alpha^2 \pi \sigma_{\rm 1}\sigma_{\rm 2}}
\label{eq-density}
\end{equation}
where $N(\leq m)$ represents the number of galaxies of which magnitudes
  are less than or equal to that of a candidate, and $\alpha=3$. 
We then obtain the LR for our sample
\begin{equation}
{\rm LR} = \frac{\alpha^2\exp(-r^{2}/2)}{2N(\leq m)}. \label{eq-LRfin}
\end{equation}
We compute LR values for all {\it IRAS}
  sources that have optical counterparts in SDSS
  using the photometric sample of galaxies. 
In order to calculate the reliability of the association using the LR values,
  we perform random associations by offsetting the positions
  of {\it IRAS} sources
  by $\approx30'$, and recompute LR values for each random
  association (e.g., \citealt{hwa07}).
Using the distribution of LR values for true and
  random associations, the reliability of each association with a
  given LR is defined by
\begin{equation}
R({\rm LR}) = 1 - \frac{N_{\rm random}({\rm LR})}{N_{\rm true}({\rm LR})}, \label{eq-rel}
\end{equation}
where $N_{\rm true}({\rm LR})$ and $N_{\rm random}({\rm LR})$ are
  the number of true and random associations with a given LR. 
In Fig. \ref{fig-lr}, we present the distribution of LR values
  and the reliability of our cross-correlation as a function of LR.
It appears that the reliability for {\it IRAS} sources decreases as LR decreases.
Therefore, we made a final sample of 13,470 IR galaxies
  by selecting {\it IRAS} sources whose reliability 
  is larger than 90\% (LR$\geq0.55$).
 
\begin{figure}
\center
\includegraphics[scale=0.85]{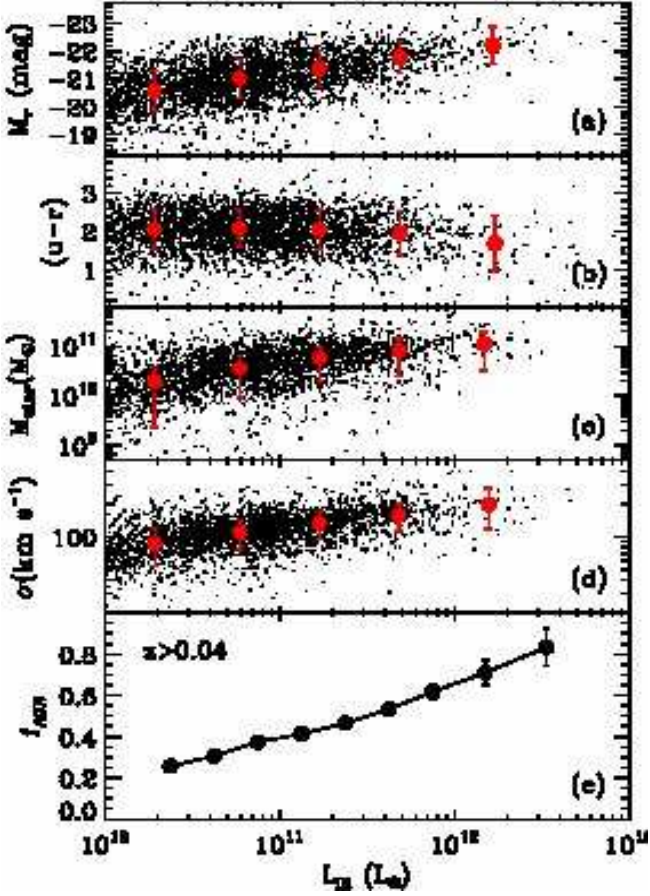}
\caption{Physical parameters of IRGs as a function of $L_{\rm IR}$.
  (a) absolute magnitude $M_r$,
  (b) ($u-r$) color, 
  (c) stellar mass $M_{\rm star}$, and
  (d) velocity dispersion $\sigma$. 
Filled circles are median values of each bin, and 
  error bars indicate the dispersion of the data.
  (e) AGN fraction (Seyferts, LINERs and composite galaxies). 
We show only 30$\%$ of IRGs in the plot, but median values and AGN fraction
  are computed using the whole IRG sample.
}\label{fig-sfrir}
\end{figure}

In the present study, we assign a single optical counterpart to a given {\it IRAS} source.
However, there can be multiple optical sources 
  within the 3 $\sigma$ positional uncertainty ellipse of the {\it IRAS} source,
  which may contribute to the measured IR flux.
This may result in the overestimation of $L_{\rm IR}$ for the optical source,
  with larger effect in high-density regions such as galaxy groups/clusters \citep{all96,floch02,bit10}.
To investigate this effect on our galaxy sample,
  we checked the ratio of the IR luminosity ($L_{\rm IR}$) to 
  the optical luminosity in $r$-band ($L_r$) as a function of the background density
  and the distance to galaxy clusters for galaxies in the volume-limited sample of L1 
  (to defined in the next section).
IR luminosity is known to correlate with 
  the optical luminosity (see Fig. \ref{fig-sfrir}a).
If we had overestimated $L_{\rm IR}$ for galaxies in high-density regions,
  the ratio $L_{\rm IR}$/$L_r$ would be expected to be artificially larger in high-density regions.
Instead, we find that this ratio does not change with the background density
  and the distance to galaxy clusters, 
  which suggests that this effect is not significant here.
The AKARI satellite is going to provide new all-sky maps with a better spatial resolution 
  than IRAS, which will definitely solve this issue \citep{mur07}.

From now on, we call galaxies with $L_{\rm IR} \geq 10^{10} L_\odot$ 
    infrared galaxies (IRGs),
  those with $10^{10} \leq L_{\rm IR} < 10^{10.5} L_\odot$ 
    lower-luminosity infrared galaxies (LLIRGs),
  those with $10^{10.5} \leq L_{\rm IR}< 10^{11} L_\odot$ 
    moderately luminous infrared galaxies (MLIRGs),  
  those with $10^{11} \leq L_{\rm IR} < 10^{12} L_\odot$ LIRGs, 
  those with $10^{12} \leq L_{\rm IR} < 10^{13} L_\odot$ ULIRGs, and 
  those with $L_{\rm IR} \geq 10^{13} L_\odot$ 
    hyperluminous infrared galaxies (HLIRGs) \citep{goto05ir}.
In the final sample, we have 11,964 IRGs including
  3049 LLIRGs, 4362 MLIRGs, 4266 LIRGs, 268 ULIRGs, and 19 HLIRGs.

Since the IRGs in this study have IR luminosities and
  $H\alpha$ emission line fluxes provided by SDSS,
  we compared the two SFR measurements for further analysis.
We convert the IR luminosity into SFR$_{\rm IR}$ 
  using the relation in \citet{ken98}: 
  SFR$_{\rm IR}$ ($M_\odot$ yr$^{-1}$) $= 1.72\times10^{-10}L_{\rm IR} (L_\odot)$.
The $H\alpha$ luminosity ($L_{H\alpha}$) is computed from the flux measurement given in
  the MPA/JHU VAGCs by applying aperture corrections
  following the method given in appendix A of \citet{hop03}.
Obscuration correction is also applied using the reddening curve of \citet{ccm89}.
Thus, SFR$_{H\alpha}$ is converted from the $L_{H\alpha}$ 
  using the relation in \citet{ken98}: 
  SFR$_{H\alpha}$ ($M_\odot$ yr$^{-1}$) 
   $= 7.9\times 10^{-42}L_{H\alpha}$ (ergs s$^{-1}$).
Fig. \ref{fig-sfr} represents the comparison between SFR$_{H\alpha}$ and SFR$_{\rm IR}$, 
  and shows that the two measurements seem to agree well
  although with a larger scatter as compared to other studies limited to local galaxies \citep{kew02,ken09}
  most probably due to the inclusion of more distant galaxies here.
It is also seen that
  SFR$_{H\alpha}$ tends to underestimate the SFR for IR luminous galaxies
  probably due to the increasing uncertainty on the obscuration correction,
  which is also seen in other studies \citep{kew02,hop03,ken09}.

\begin{figure}
\center
\includegraphics[scale=0.8]{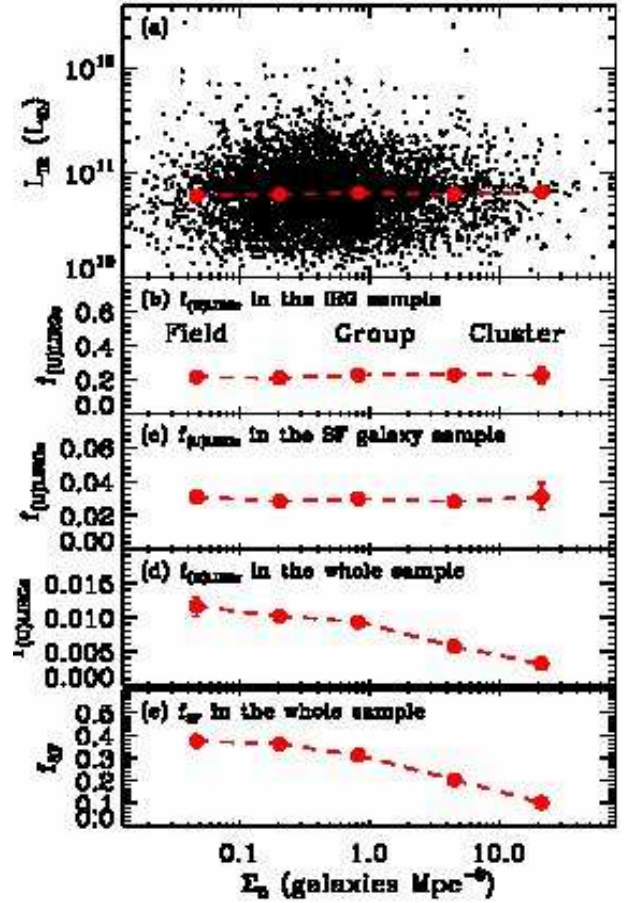}
\caption{(a) $L_{\rm IR}$ of IRGs in L1 as a function of $\Sigma_5$.
Filled circles are median values at each density bin.
(b) Fraction of LIRGs plus ULIRGs among IRGs
  as a function of $\Sigma_5$.
(c) Fraction of LIRGs plus ULIRGs among SF galaxies in L1
  (SFR$_{\rm H\alpha}\geq1$ or SFR$_{\rm IR}\geq1$ M$_\odot$yr$^{-1}$). 
(d) Fraction of LIRGs plus ULIRGs among all galaxies in L1.
(e) Fraction of SF galaxies among all galaxies in L1.
}\label{fig-irsig1d}
\end{figure}

\subsection{Galaxy Environment}\label{env}

We define a volume-limited sample of galaxies selected by absolute magnitude 
  and redshift limits as shown in Fig. \ref{fig-vol}(c).
We define two samples, $L1$ and $L2$,
  that were chosen to maximize the number of target IRGs 
  (to be discussed in the end of this section)
  and the number of ULIRGs in the volume, respectively:
  L1 ($0.03\leq z<0.0788$ and $M_r\leq -19.91$) having 134,272 galaxies
  including 5513 IRGs (6 ULIRGs) and
  L2 ($0.03\leq z<0.17$ and $M_r\leq -21.68$) having 129,851 galaxies
  including 2157 IRGs (38 ULIRGs).
The apparent magnitude limit line ($m_r=17.77$) for the SDSS main galaxy sample
  shown in Fig. \ref{fig-vol}(c) 
  is obtained using the mean $K$-correction relation given by eq. (2) of \citet{choi07}.

We consider three kinds of environment indicators:
  a surface galaxy number density estimated from 
    the five nearest neighbor galaxy ($\Sigma_5$)
    as a large-scale environmental parameter,
  a distance to the nearest neighbor galaxy ($R_{\rm n}$)
    as a small-scale environmental parameter, and 
  a distance to a galaxy cluster ($R$).

\begin{figure*}
\center
\includegraphics[scale=0.8]{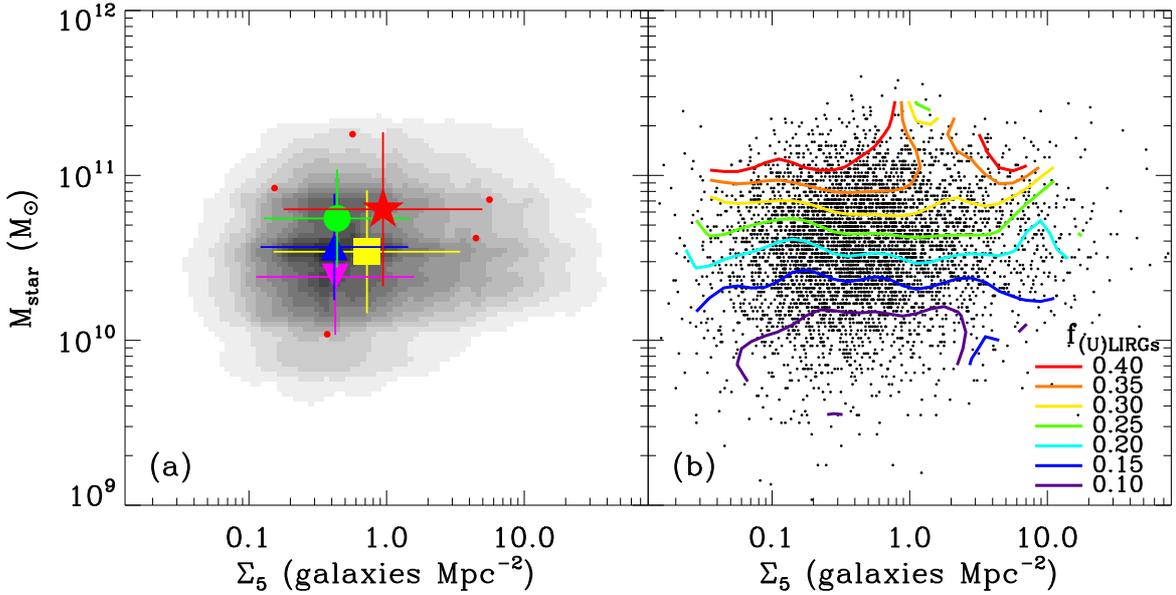}
\caption{(a) Distributions of IRGs in L1 in the space of $M_{\rm star}$ and $\Sigma_5$.
ULIRGs, LIRGs, MLIRGs, LLIRGs, and non-SF galaxies are plotted with 
  star, circle, triangle, upside down triangle, and square, respectively.
Non-SF galaxies are those with SFR$_{\rm H\alpha}=0$ and SFR$_{\rm IR}=0$ ($M_\odot$ yr$^{-1}$).
Error bars indicate the $1\sigma$-dispersion of the distribution.
Grey contour map is the number density of non-SF galaxies, and red dots are ULIRGs.
(b) $f_{\rm (U)LIRGs}$ contours in the space of $M_{\rm star}$ and $\Sigma_5$.
Dots are IRGs.
}\label{fig-irsig2d}
\end{figure*}

The background density, $\Sigma_5$, is defined by $\Sigma_5=5(\pi D^2_{p,5})^{-1}$.
$D_{p,5}$ is the projected distance to the 5th-nearest neighbor.
The 5th-nearest neighbor of each target galaxy is identified 
  in the volume-limited sample of L1 among the neighbor galaxies 
  that have velocities relative to the target galaxy 
  less than $1500$ km s$^{-1}$ to exclude foreground and background galaxies.
The $\Sigma_5$ for L2 galaxies is computed 
  using the same method as for L1 galaxies.

To investigate the environment of galaxy clusters in detail, 
  we compute the distance to galaxy clusters as a second environmental parameter
  using the method given in \citet{ph09}.
We use the Abell clusters and the sample of 929,234 galaxies in this study. 
The radius of $r_{200,{\rm cl}}$ (usually called the virial radius)
  for each cluster where the mean overdensity drops to 200 times 
  the critical density of the universe $\rho_{\rm c}$
  is computed using the formula given by \citet{car97}:
\begin{equation}
r_{200,{\rm cl}}= \frac{3^{1/2}\sigma_{\rm cl}}{10 H(z)},
\end{equation}
where $\sigma_{\rm cl}$ is the velocity dispersion of a cluster and
  the Hubble parameter at $z$ is
  $H^2(z)=H^2_0 [\Omega_m(1+z)^3 +\Omega_k(1+z)^2+\Omega_\Lambda]$ \citep{pee93}.
$\Omega_m$, $\Omega_k$, and $\Omega_\Lambda$ are the dimensionless density parameters.

To define the small-scale environmental parameter attributed to the nearest neighbor,
  we first find the nearest neighbor of a target galaxy
  that is the closest to the target galaxy on the projected sky
  and satisfies the conditions of magnitude and relative velocity.
We searched for the nearest neighbor galaxy among galaxies
  that have magnitudes brighter than $M_r=M_{r,\rm target}+0.5$ and 
       have relative velocities less than 
       $\Delta \upsilon=|\upsilon_{\rm neighbors}- \upsilon_{\rm target}|=600$ km s$^{-1}$
       for early-type target galaxies and less than $\Delta \upsilon=400$ km s$^{-1}$ 
       for late-type target galaxies.
These velocity limits cover most close neighbors as seen in Fig. 1 of \citet{park08}.
Since we use a volume-limited sample of galaxies with $M_r\leq-19.91$ (for L1)
  and $M_r\leq -21.68$ (for L2),
  we often restrict our analysis to target galaxies 
  brighter than $M_{r,\rm target}=-20.41$ (for L1) and
  $M_{r,\rm target}=-22.18$ (for L2) so that their neighbors are complete
  when we investigate the effects of the nearest neighbor galaxy.

\begin{figure}
\center
\includegraphics[scale=0.8]{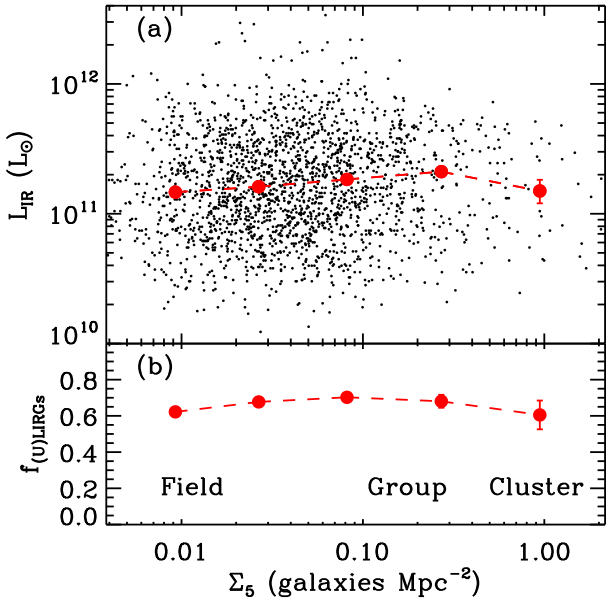}
\caption{(a) $L_{\rm IR}$ of IRGs in L2 as a function of $\Sigma_5$.
Filled circles are median values at each density bin.
(b) Fraction of LIRGs plus ULIRGs among IRGs
  as a function of $\Sigma_5$ in L2.
 }\label{fig-irsig1dl2}
\end{figure}

\begin{figure}
\center
\includegraphics[scale=0.8]{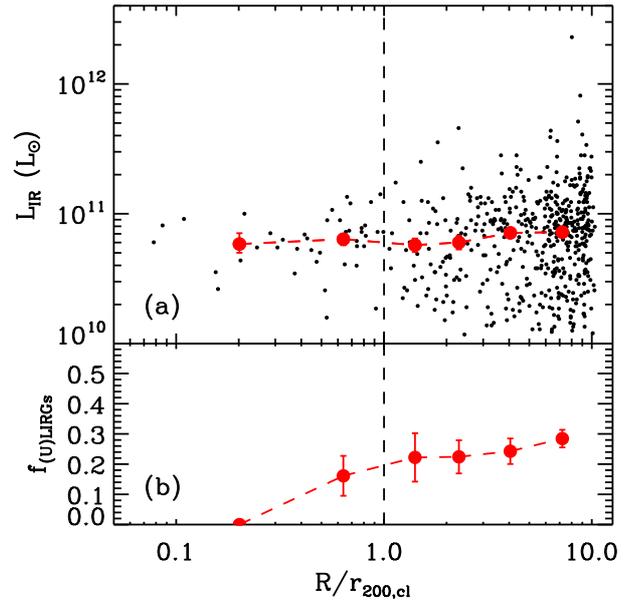}
\caption{(a) $L_{\rm IR}$ of IRGs in L1
  as a function of the clustercentric radius normalized to 
  the cluster virial radius ($R/r_{\rm 200,cl}$).
Filled circle is the median value of each density bin.
(b) $f_{\rm (U)LIRGs}$ vs. $R/r_{\rm 200,cl}$.
}\label{fig-ircl1d}
\end{figure}

\begin{figure*}
\center
\includegraphics[scale=0.8]{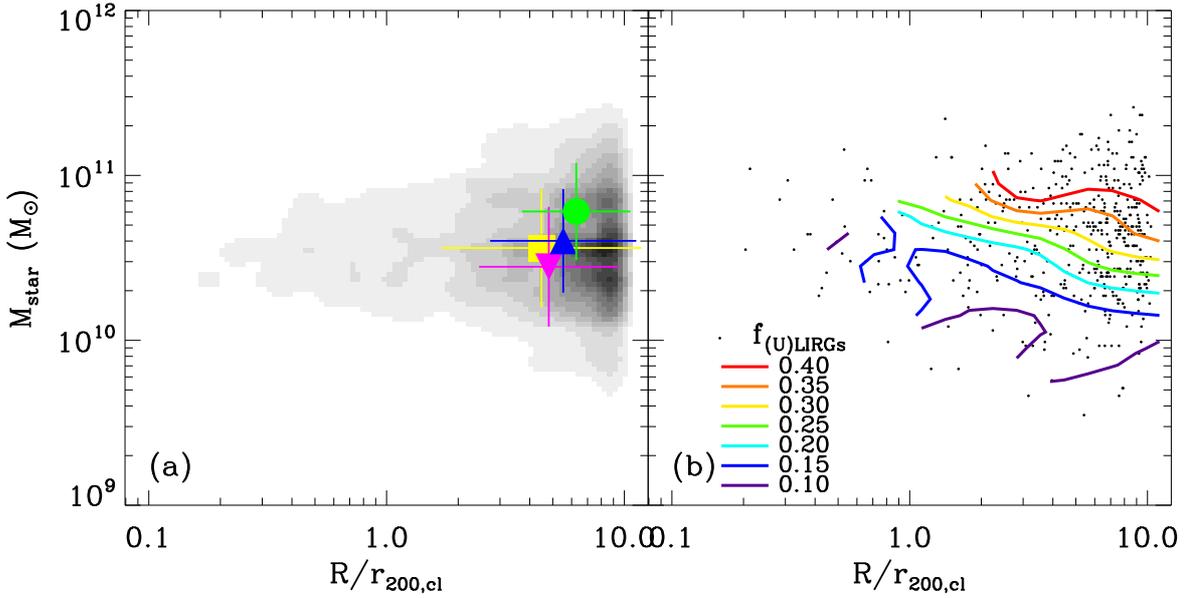}
\caption{(a) Distributions of IRGs in L1
  in the space of the stellar mass and the clustercentric radius.
Symbols are same as in Fig. \ref{fig-irsig2d}.
(b) $f_{\rm (U)LIRGs}$ contours
  in the space of the stellar mass and the clustercentric radius.
  Dots are IRGs.
}\label{fig-ircl2d}
\end{figure*}

\begin{figure}
\center
\includegraphics[scale=0.8]{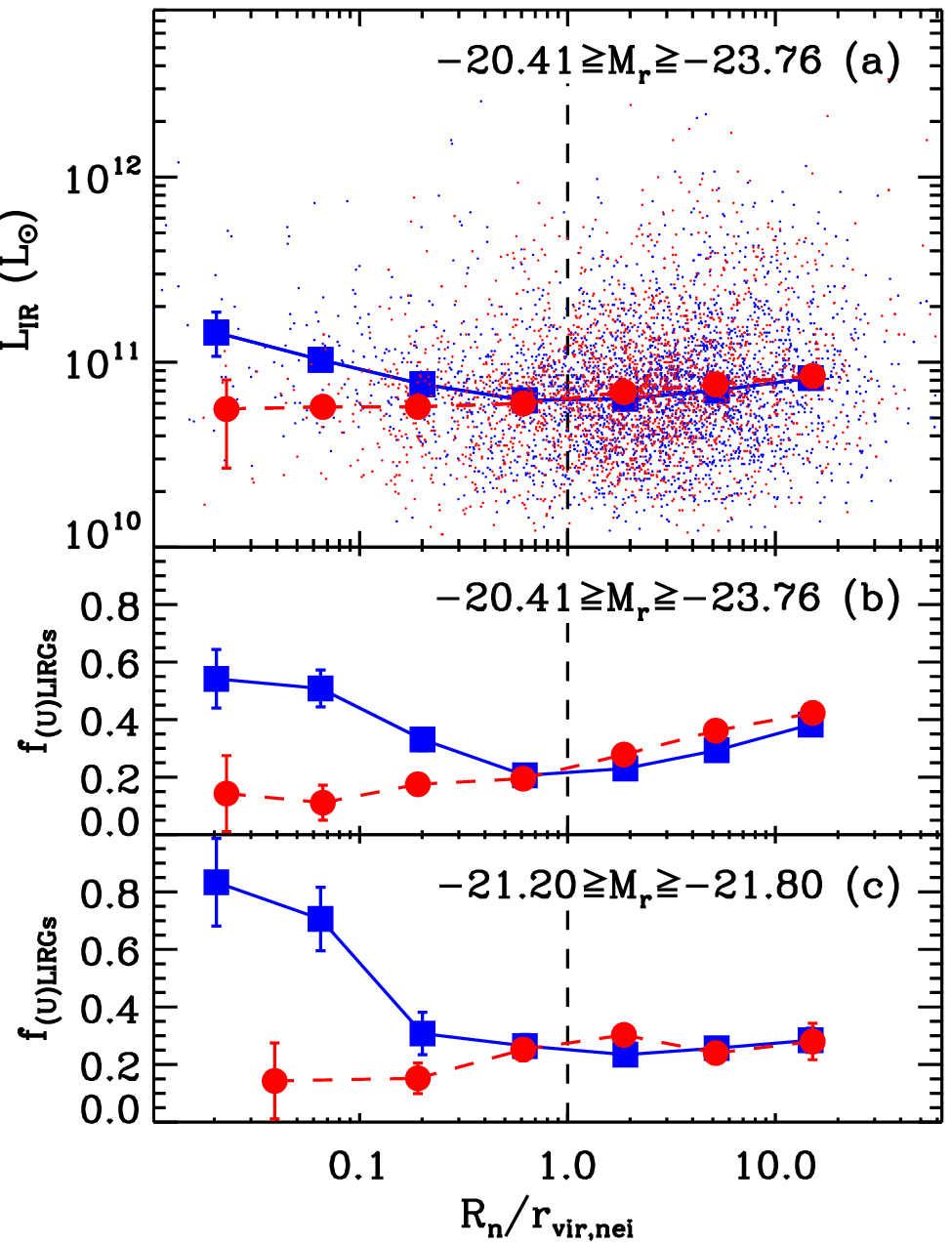}
\caption{(a) $L_{\rm IR}$ of IRGs in L1 plus L2 ($-20.41\geq M_r\geq-23.76$)
 as a function of the projected distance 
  to the nearest neighbor galaxy ($R_{\rm n}/r_{\rm vir,nei}$).
Red and blue dots are the galaxies having early- and late-type 
  nearest neighbor galaxies, respectively. 
Circle (early-type neighbor case) and square (late-type neighbor case)
  are median values of each density bin. 
$f_{\rm (U)LIRGs}$ vs. $R_{\rm n}/r_{\rm vir,n}$ 
  (b) for IRGs in L1 plus L2 ($-20.41\geq M_r\geq-23.76$) and 
  (c) for IRGs with $-21.20\geq M_r\geq-21.80$ among L1 plus L2.
}\label{fig-irneib1d}
\end{figure}

The virial radius of a galaxy within which 
  the mean mass density is 200 times the critical density 
  of the universe ($\rho_c$) is computed by
\begin{equation}
r_{\rm vir}=(3 \gamma L/4\pi)^{1/3} (200\rho_c)^{-1/3},
\label{eq-vir}
\end{equation}

where $L$ is the galaxy luminosity, and $\gamma$ is the mass-to-light ratio.
We assume that the mass-to-light ratio of early-type galaxies
  is on average twice as large as that of late-type galaxies
  at the same absolute magnitude $M_r$,
  which means $\gamma$(early)$=2\gamma$(late) 
  [see \S 2.5 of \citet{pc09} and \S 2 of \citet{park08}]. 

Since we adopt $\Omega_m = 0.27$, $200\rho_c = 200 {\bar\rho}/\Omega_m = 740{\bar\rho}$
  where $\bar\rho$ is the mean density of the universe.
The value of mean density of the universe, 
  $\bar\rho=(0.0223\pm0.0005)(\gamma L)_{-20} (h^{-1}{\rm Mpc})^{-3}$,
  was adopted where $(\gamma L)_{-20}$ is the mass of a late-type galaxy 
  with $M_r=-20$ \citep{park08}.
According to our formula the virial radii of galaxies with
  $M_r=-19.5,-20.0,$ and $-20.5$ are 260, 300, and 350 $h^{-1}$ kpc for early types,
  and 210, 240, and 280 $h^{-1}$ kpc for late types, respectively.  

\section{Results}\label{results}
\subsection{Physical Parameters as a function of IR luminosity}\label{parameter}

We show several physical parameters as a function of 
  IR luminosity for IRGs in Fig. \ref{fig-sfrir}.
In this plot, we use all the IRGs instead of the volume-limited sample of galaxies. 
It is seen that, as $L_{\rm IR}$ increases, 
  IRGs become more luminous in $r$-band (a), bluer in ($u-r$) color (b)
  and more massive (c, d).
Interestingly, bluer colors for more IR luminous galaxies
  may be incompatible with the idea that IR luminous galaxies
  are dust-enshrouded systems.
However, this can be explained by the idea that 
  the optical colors of ULIRGs
  are dominated by the extended distribution of young, blue stars, and
  the dust extinction is patchy \citep{chen10}.
\citet{vei99} also found bluer optical colors for more IR luminous galaxies
  in the sample of ULIRGs,
  and suggested that the optical continuum was less affected by
  dust extinction than the emission-line gas near hot stars or AGNs.
  
We measure the fraction of galaxies showing AGN activity 
  among emission-line galaxies in (e).
We determine the spectral types of emission-line galaxies 
  based on the criteria given by \citet{kew06} using the emission line ratio diagram, 
  commonly known as Baldwin-Phillips-Terlevich (BPT) diagram \citep{bpt81}: 
  star-forming galaxies, Seyferts, LINERs, composite galaxies, and ambiguous galaxies.
We regard Seyferts plus LINERs and composite galaxies
  as galaxies showing AGN activity,
  and restrict our analysis to the galaxies at $z>0.04$ 
  due to the problem of small (3\arcsec) fixed-size aperture \citep{kew06}.
The uncertainties of the fraction represent $68\%$ $(1\sigma)$ confidence intervals 
  that are determined by the bootstrap resampling method.
Interestingly,
  the AGN fraction for IRGs keeps increasing as $L_{\rm IR}$ increases
  over all the IR luminosity range,
  which is consistent with previous results (e.g., \citealt{vei95,vei02,yuan10}).

We wish to emphasize that the sample of (U)LIRGs used in this study
  is dominated by LIRGs (99.5\% and 97.4\% among (U)LIRGs in L1 and L2, respectively),
  thus the results in next sections are mostly determined by LIRGs 
  (to be discussed in \S\ref{power}).

\subsection{Background Density Dependence of IRG Properties}\label{sigma}

In Fig. \ref{fig-irsig1d},
  we present the IR luminosity of IRGs and the fraction of LIRGs plus ULIRGs
  among several galaxy samples ($f_{\rm (U)LIRGs}$) in L1
  as a function of the background density ($\Sigma_5$).

It appears that neither the median $L_{\rm IR}$ values (Fig. \ref{fig-irsig1d}a) 
  nor $f_{\rm (U)LIRGs}$ (in the IRG sample, b) 
  change with $\Sigma_5$.
Note that $f_{\rm (U)LIRGs}$ in (b) is the fraction of LIRGs plus ULIRGs 
  among IRGs, which is different from the fraction of 
  SF galaxies that are usually found to decrease monotonously
  as the background density increases. 
This is due to the large contribution of non-SF galaxies in high-density regions (see e).
When we compute $f_{\rm (U)LIRGs}$ in the whole sample
  regardless of IR and $H\alpha$ 
  detections (Fig. \ref{fig-irsig1d}d), it decreases as $\Sigma_5$ increases.
On the other hand, if we compute $f_{\rm (U)LIRGs}$ in the SF galaxy sample
  as shown in (c),
  we find a similar trend of $f_{\rm (U)LIRGs}$ to the case of IRGs in (b).
We call SF galaxies those with SFR$_{\rm IR}$ or 
  SFR$_{\rm H\alpha}\geq1$ M$_\odot$yr$^{-1}$.
We include galaxies having $H\alpha$ measurements
  in order to take into account for SF galaxies that are not detected in IR bands
  due to the IR detection limit.  
Since IRGs can be regarded as a representative sample of SF galaxies 
 (IRGs are by definition galaxies with SFR$_{\rm IR}>1.72$ M$_\odot$yr$^{-1}$
  in this study) and we are interested in the triggering mechanism of (U)LIRGs 
  that are extreme cases among IRGs,
  we are going to focus on $f_{\rm (U)LIRGs}$ in the IRG sample
  in further analysis.
When we use only the sample of IRGs hosting AGNs in (b),
  no dependence of $f_{\rm (U)LIRGs}$ on the background density is seen again.

Since the SFA of galaxies is strongly linked to the stellar mass
  \citep{bri04,noe07sf,noe07mass,elb07},
  it is important to quantify its effect in order to
  investigate the effects of environment.
Thus we plot the distribution of IRGs in company with the non-SF galaxies 
  in two-dimensional space of the stellar mass ($M_{\rm star}$) and 
  the background density ($\Sigma_5$) in Fig. \ref{fig-irsig2d}(a).
Non-SF galaxies are those having neither $H\alpha$ line emissions nor IR emission
  (SFR$_{\rm H\alpha}=0$ and SFR$_{\rm IR}=0$ $M_\odot$ yr$^{-1}$).
It is clearly seen that the stellar mass of IRGs increases as $L_{\rm IR}$ increase,
  while the median values of $\Sigma_5$ do not change much with $L_{\rm IR}$.
IRGs tend to be found in low-density regions compared to non-SF galaxies.   
Panel (b) shows constant $f_{\rm (U)LIRGs}$ contours
   in the same parameter space.
Interestingly, the $f_{\rm (U)LIRGs}$ contours are nearly horizontal,
  which confirms that the dependence of $f_{\rm (U)LIRGs}$ on the background density 
  is very weak compared to that on the stellar mass. 
When we use galaxies in L2 that contain more ULIRGs than L1,
  no dependence of $f_{\rm (U)LIRGs}$ is similarly found again (see Fig. \ref{fig-irsig1dl2}).

\subsection{Effect of Galaxy Clusters on IRG Properties}\label{cluster}

In Fig. \ref{fig-ircl1d},
  we present $L_{\rm IR}$ of IRGs and $f_{\rm (U)LIRGs}$
  as a function of the projected distance to the galaxy cluster ($R$).
We restrict this plot to the IRGs whose clustercentric distances are measured
  in this study.
Fig. \ref{fig-ircl1d}(a) shows that the median $L_{\rm IR}$ values do not change 
  much with the clustercentric distance.
In panel (b), $f_{\rm (U)LIRGs}$ appears to decrease with large error-bars,
  as we approach the center of galaxy clusters.
Moreover, in the very center of galaxy cluster ($R<0.5r_{\rm 200,cl}$),
  no (U)LIRGs are found.

We plot, in Fig. \ref{fig-ircl2d}, the distribution of IRGs
  in the space of the stellar mass and the clustercentric radius.
The right panel shows that $f_{\rm (U)LIRGs}$ decreases as IRGs are located closer
  to the center of clusters.
The dependence of $f_{\rm (U)LIRGs}$ on the clustercentric radius
  is much weaker than the dependence on the stellar mass.
According to our finding in the next section this dependence can appear
  because the typical morphology of the nearest neighbor changes from
  late type to early type.
The IRGs near the central region are typically under the strong influence of
  early-type neighbors, and are difficult to show (U)LIRGs behaviors.

\subsection{Nearest Neighbor Dependence of IRG Properties}\label{neighb}

To measure the effect of the nearest neighbor galaxy on the SFA of IRGs,
  we plot $L_{\rm IR}$ of IRGs and $f_{\rm (U)LIRGs}$
  as a function of the distance to the nearest neighbor in Fig. \ref{fig-irneib1d}.
In order to remain complete in terms of neighbor galaxies, we restrict our analysis
  to the galaxies having $M_r\leq -20.41$ and 
  $M_r\leq -22.18$ in L1 and L2, respectively.
Panels (a-b) show that $L_{\rm IR}$ and $f_{\rm (U)LIRGs}$ depend on 
  the projected distance to the 
  nearest neighbor as well as on the neighbor's morphology.
When an IRG is located farther than 0.5 $r_{\rm vir,nei}$ 
  (virial radius of neighbor galaxy),
  $f_{\rm (U)LIRGs}$ decreases as the distance to the neighbor decreases,
  but its dependence on the neighbor's morphology is negligible.
On the other hand, when an IRG is located at $R_{\rm n} \lesssim 0.5 r_{\rm vir,nei}$,
  $f_{\rm (U)LIRGs}$ increases as the target IRG approaches a late-type neighbor, but
  decreases as it approaches an early-type neighbor.
It is important to note that the bifurcation of $f_{\rm (U)LIRGs}$
  depending on the neighbor's morphology
  occurs at $R_{\rm n}\approx 0.5 r_{\rm vir,nei}$,
  which is similarly seen in the plot of median $L_{\rm IR}$ values along with 
  the distance to the nearest neighbor.

In order to check whether the difference in $f_{\rm (U)LIRGs}$ between 
  IRGs with a late- and early-type close neighbor is real or 
  if it could be explained by small number statistics, 
  we made the following experiment.
We produced 1000 Monte-Carlo realizations of the same number of galaxies
  with close neighbors by randomly selecting galaxies from the larger 
  sample of galaxies with no neighbor closer than $R_{\rm n}=r_{\rm vir,nei}$
  (24 and 7 galaxies with late- and early-type neighbors 
  at $R_{\rm n}<0.03 r_{\rm vir,nei}$, respectively). 
While in the real dataset, $f_{\rm (U)LIRGs}$ is 3.8 times larger for IRGs 
  with a late-type neighbor than for those with an early-type neighbor, 
  such large difference is found in only 6\% of the random samples
  which suggests that the difference is real.
This experiment supports the robustness of the bifurcation of $f_{\rm (U)LIRGs}$
  depending on the neighbor's morphology at $R_{\rm n}\lesssim 0.5 r_{\rm vir,nei}$.
  
The trends seen in panel (b) of Fig. \ref{fig-irneib1d} with respect to $R_{\rm n}$
  are actually coupled with luminosity dependence of $f_{\rm (U)LIRGs}$.
To disentangle $f_{\rm (U)LIRGs}$ from the coupling with luminosity and
  to see its dependence on the neighbor environment only,
  we fix the absolute magnitude of IRGs to a narrow range of $-21.20\geq M_r\geq-21.80$ 
  in panel (c) of  Fig. \ref{fig-irneib1d}.
Now $f_{\rm (U)LIRGs}$ remains nearly constant down to $R_{\rm n} \approx 0.5 r_{\rm vir,nei}$,
  which indicates no direct neighbor effect at these separations.
The bifurcation of $f_{\rm (U)LIRGs}$ still occurs at $R_{\rm n} \approx 0.5 r_{\rm vir,nei}$.
We will discuss these findings in more detail in \S\ref{power}.

\begin{figure*}
\center
\includegraphics[scale=0.8]{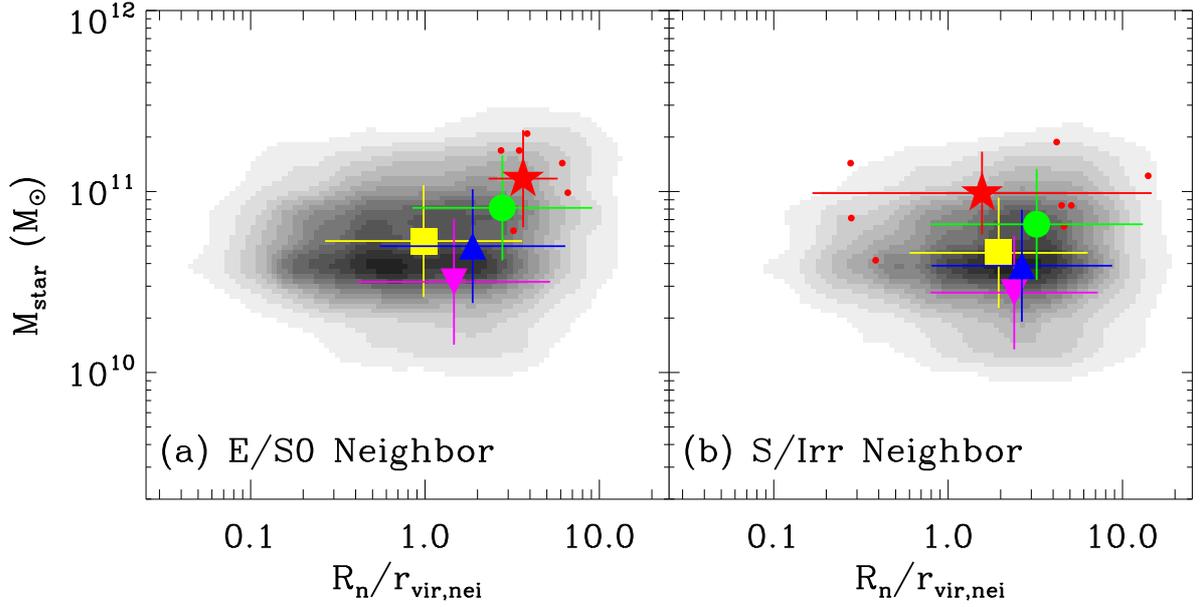}
\caption{Distributions of IRGs in L1 plus L2 in the space of 
  the stellar mass and the distance to the nearest neighbor galaxy
  when the nearest galaxy is an early type (a) and a late type (b). 
Symbols are same as in Fig. \ref{fig-irsig2d}.
Red dots are ULIRGs.
}\label{fig-irneib2da}
\end{figure*}

\begin{figure*}
\center
\includegraphics[scale=0.8]{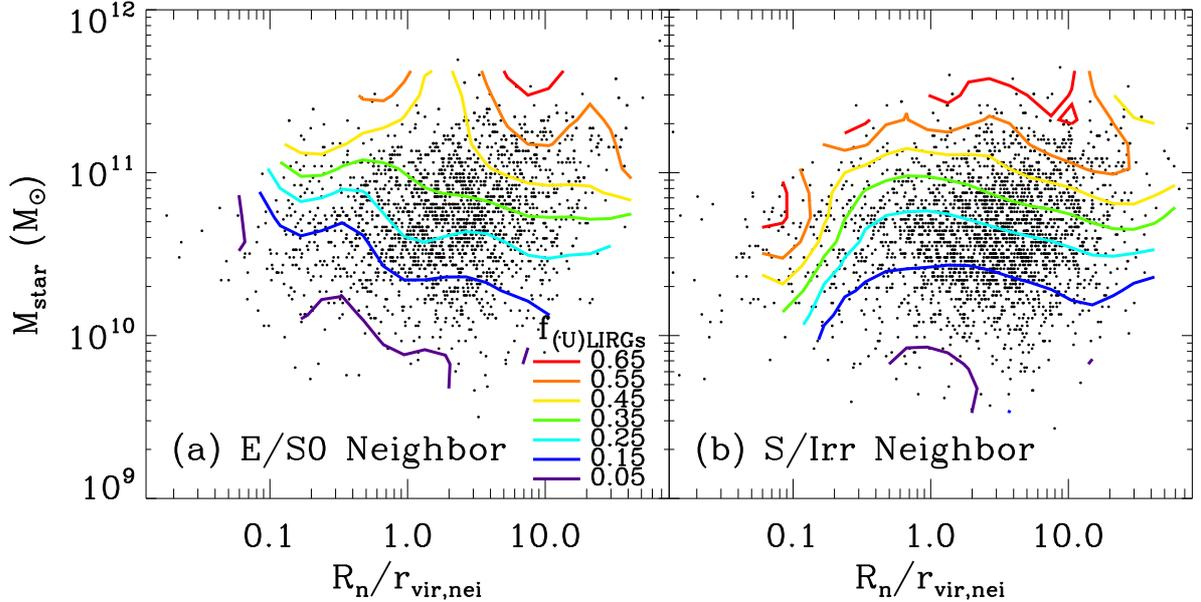}
\caption{$f_{\rm (U)LIRGs}$ for IRGs in in L1 plus L2 in the space of $M_{\rm star}$ and 
  $R_{\rm n}/r_{\rm vir,nei}$
  when the nearest galaxy is an early type (a) and a late type (b). 
Dots are IRGs.
}\label{fig-irneib2db}
\end{figure*}

In Fig. \ref{fig-irneib2da}, we present the distribution of IRGs 
  in the space of the stellar mass and the distance to the nearest neighbor.
When the nearest neighbor galaxy of IRGs is an early type (a),
  IRGs are preferentially located farther away from their neighbor than non-SF galaxies,
  and more luminous IRGs are farther away from their neighbor. 
On the other hand,
  the median distance to the nearest neighbor for the late-type neighbor case (b)
  is not clearly distinguishable among IRGs including non-SF galaxies.

\begin{figure}
\center
\includegraphics[scale=0.8]{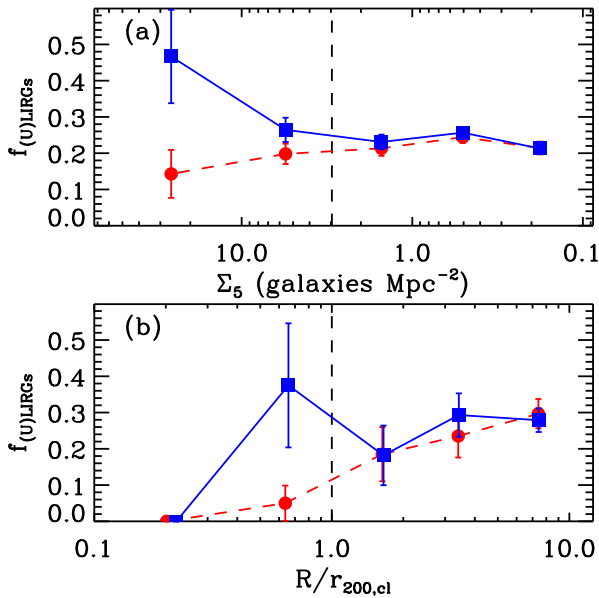}
\caption{(a) $f_{\rm (U)LIRGs}$ for IRGs in L1 as a function of $\Sigma_5$.
Circle (early-type neighbor case) and square (late-type neighbor case)
  are median values of each density bin. 
Vertical dashed line indicates the $\Sigma_5$ that roughly corresponds to
  one virial radius of the clustercentric radius ($R\sim r_{\rm 200,cl}$)
  determined from the comparison between 
  $\Sigma_5$ and the distance to galaxy clusters in L1.
(b) Same as above, but for $f_{\rm (U)LIRGs}$ vs. $R/r_{\rm 200,cl}$.
}\label{fig-ir1dneib}
\end{figure}

The $f_{\rm (U)LIRGs}$ contours in the same parameter space
  are plotted in Fig. \ref{fig-irneib2db} depending on the 
  morphology of the nearest neighbor. 
Unlike the $f_{\rm (U)LIRGs}$ contours in the space of $M_{\rm star}$ and $\Sigma_5$
  seen in Fig. \ref{fig-irsig2d}(b),
  the constant $f_{\rm (U)LIRGs}$ contours in Fig. \ref{fig-irneib2db}
  depend on both $M_{\rm star}$ and $R_{\rm n}$.
When an IRG is located at $R_{\rm n}\gtrapprox 0.5 r_{\rm vir,nei}$,
  the probability for IRGs to become (U)LIRGs increases
  as IRGs are more massive and are farther away from their neighbor
  regardless of the morphology of the nearest neighbor.
However, if an IRG is located at $R_{\rm n}\lessapprox 0.5 r_{\rm vir,nei}$,
  the dependence of $f_{\rm (U)LIRGs}$ on $M_{\rm star}$ and $R_{\rm n}$
  is significantly different depending on the morphology of the nearest neighbor.
For the early-type neighbor case, 
  $f_{\rm (U)LIRGs}$ increases as IRGs are more massive and are farther away 
  from their neighbor, while
  $f_{\rm (U)LIRGs}$ increases as IRGs are more massive and are closer to 
  their neighbor for the late-type neighbor case.
The different behavior of $f_{\rm (U)LIRGs}$ depending on 
  the morphology of the nearest neighbor indicates an important
  role of the nearest neighbor on the activity of IRGs.

Bearing in mind that there is a strong effect of the nearest neighbor galaxy on $f_{\rm (U)LIRGs}$,
  we re-investigate the dependence of $f_{\rm (U)LIRGs}$ on the background density
  and the distance to galaxy clusters.
Fig. \ref{fig-ir1dneib} shows $f_{\rm (U)LIRGs}$ as a function of the background density
  and the distance to galaxy clusters 
  as seen in Figs. \ref{fig-irsig1d}, \ref{fig-irsig1dl2} and \ref{fig-ircl1d},
  but depending on the morphology of nearest neighbor galaxy.
It is seen that $f_{\rm (U)LIRGs}$ for the late-type neighbor case, on average, are larger 
  than those for early-type neighbor case.
The difference of $f_{\rm (U)LIRGs}$ are insignificant in low-density regions
  where the distance to the nearest neighbor is large.
However, the difference of $f_{\rm (U)LIRGs}$ becomes prominent in high-density regions
  (even in the cluster region of $R\sim0.6r_{\rm 200,cl}$)
  where the neighbor separation is smaller compared to low-density regions
  so that IRGs experience the effect of the neighbor galaxy. 
These confirm again the strong effects of the nearest neighbor on $f_{\rm (U)LIRGs}$.

\section{Discussions}\label{discuss}
\subsection{What powers LIRGs and ULIRGs?}\label{power}

The strong dependence of $f_{\rm (U)LIRGs}$ and median $L_{\rm IR}$ values
  on the morphology of and the distance to the nearest neighbor galaxy 
  seen in \S \ref{neighb} supports the idea that LIRGs and ULIRGs are triggered
  by galaxy-galaxy interactions/merging.
In particular, the bifurcation of $f_{\rm (U)LIRGs}$ and median $L_{\rm IR}$ values
  occurs at $R_{\rm n}\approx 0.5 r_{\rm vir,nei}$
  where the effects of galaxy-galaxy interactions can significantly appear \citep{pc09}.
The bifurcation of $f_{\rm (U)LIRGs}$ and median $L_{\rm IR}$ values
  at $R_{\rm n}\approx 0.5 r_{\rm vir,nei}$ depending on the morphology
  of the neighbor may imply that the hydrodynamic interactions
  with the nearest neighbor play a critical role in triggering SFA of IRGs
  in addition to the tidal interactions.
If an IRG approaches a late-type neighbor 
  within half of the virial radius of the neighbor,
  the inflow of cold gas from the neighbor into the target IRG increases and
  the SFA of IRGs can be boosted.
On the other hand, if an IRG approaches an early-type neighbor 
  within the virial radius of the neighbor,
  the hot gas of the early-type neighbor prevents 
  the IRG from forming stars with cold gas 
  or there is no inflow of cold gas from its early-type neighbor, 
  so the SFA of IRG is not boosted even if it has a close companion.
The SF quenching mechanisms of hot gas in early-type neighbors
  could be similar to those of hot intracluster medium of galaxy clusters
  acting on late-type galaxies in it,
  which are hydrodynamic processes such as 
  thermal evaporation, strangulation, ram pressure stripping, or
  viscous stripping \citep{pc09,ph09}.
IRGs may not necessarily be late types.
In fact, 4\% of IRGs in this study are early types,
  which shows again the importance of hydrodynamic interactions 
  with late-type neighbors for the SFA.

\begin{figure}
\center
\includegraphics[scale=0.8]{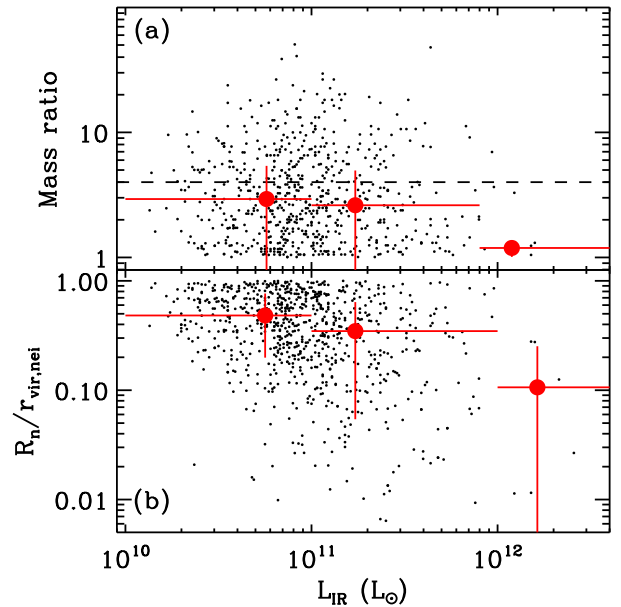}
\caption{(a) Stellar mass ratio between IRGs and 
  their nearest neighbor galaxy, and
 (b) projected distance to the nearest neighbor galaxy
  of IRGs ($R_{\rm n}/r_{\rm vir,nei}$) 
  as a function of $L_{\rm IR}$ of IRGs in L1 plus L2.
IRGs ($M_r\leq-21$) having late-type neighbors at $R_{\rm n}\leq r_{\rm vir,nei}$
  are shown. 
Large filled circles and vertical error-bars represent the median values
  and their errors in $L_{\rm IR}$ bins
  that are represented by horizontal error-bars .
The dashed horizontal line in (a) divides IRGs into major (mass ratio$<4$)
  and minor (mass ratio$>4$) mergers.
 }\label{fig-ratio}
\end{figure}

\begin{figure*}
\center
\includegraphics[scale=0.8]{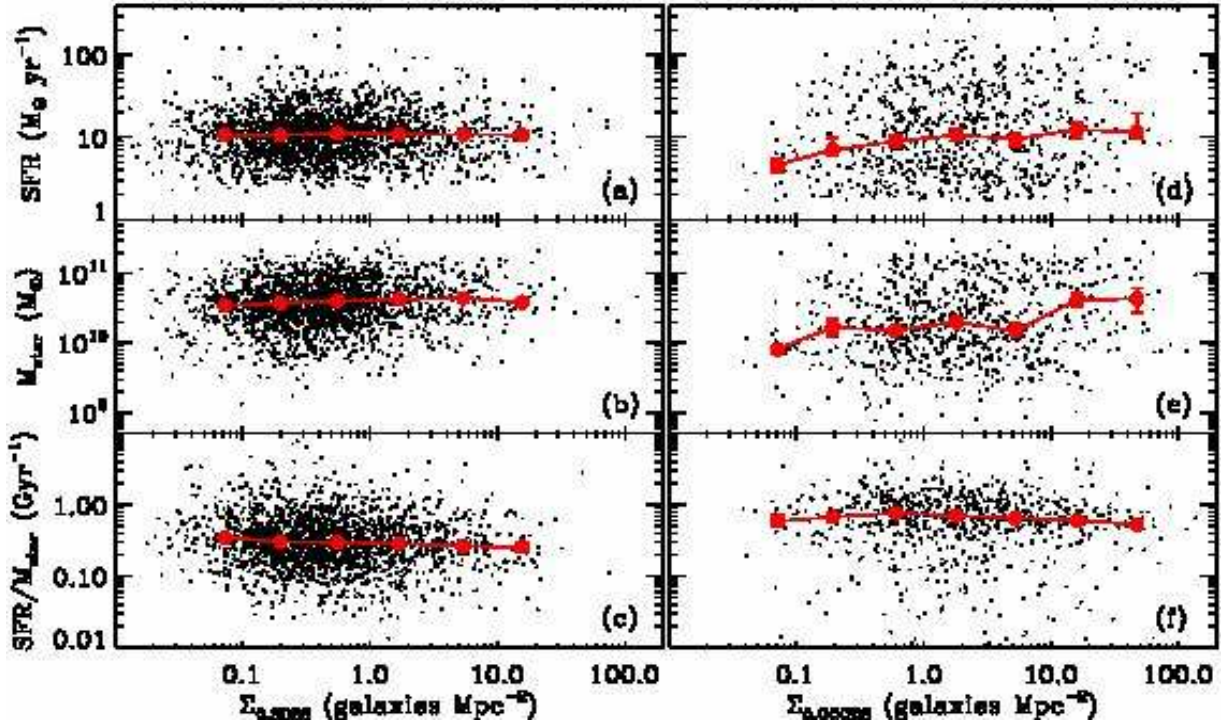}
\caption{(a) SFRs of IRGs in L1,
  (b) stellar mass of IRGs, $M_{\rm star}$, and
  (c) specific SFRs of IRGs as a function of $\Sigma_5$.
(d-f) Same as (a-c), but for the high-redshift IRGs ($0.8<z<1.2$) studied in \citet{elb07}.
Median value and its error at each bin is shown by filled circle and its error-bar.
}\label{fig-sfrenv1}
\end{figure*}

Interestingly, if we see Fig. \ref{fig-irneib1d} (a-b),
  $f_{\rm (U)LIRGs}$ and median $L_{\rm IR}$ values 
  at the largest neighbor separations ($R_{\rm n} \sim 15 r_{\rm vir,nei}$) 
  appear to be larger than those at the intermediate neighbor separation 
  ($R_{\rm n} \sim 2 r_{\rm vir,nei}$),
    which may provide another hint for the evolution of (U)LIRGs.
As an IRG experiences interactions/merging with its closest neighbor galaxy,
  the galaxies in pair tend to be (U)LIRGs though the probability 
  to become (U)LIRGs is different depending on the morphology of the neighbor.
When two galaxies just finished to merge,
  the end product of the merger will be bright due to the very recent SFA 
  and/or the increase of mass,
  and the new nearest neighbor galaxy of the merger product will be far away.
This seems to be related to the fact that some (U)LIRGs are in pair
  and others are found as a single system (e.g., \citealt{dckim02,vei02}),
  and to the variation of SFA of (U)LIRGs along with the merger stage (e.g., \citealt{mur01}).
The SFA then would decrease until the galaxy
  experiences new interactions/merging with the nearest neighbor galaxy.
This scenario is consistent with the fact that 
  there is no dependence of $f_{\rm (U)LIRGs}$ and median $L_{\rm IR}$ values 
  on the neighbor's morphology at this separation ($R_{\rm n}>0.5r_{\rm vir,nei}$),
  and could explain why $f_{\rm (U)LIRGs}$ and median $L_{\rm IR}$ values
  at the largest neighbor separation are
  larger than those at the intermediate neighbor separation.
As we show in Fig. \ref{fig-irneib1d}(c),
  if we fix the absolute magnitude of IRGs (or stellar mass),
  the bifurcation of $f_{\rm (U)LIRGs}$ depending on the morphology
  of the nearest neighbor at $R_{\rm n} \lesssim 0.5 r_{\rm vir,nei}$ is still seen,
  but $f_{\rm (U)LIRGs}$ (and median $L_{\rm IR}$ values) 
  remains constant at large separations ($R_{\rm n}>0.5r_{\rm vir,nei}$),
  which is consistent with the interpretation above.
These results seem to be compatible with recent studies 
  focusing on the effects of the nearest neighbor galaxies on the evolution of 
  ``normal'' galaxies \citep{park08,pc09,ph09,hp09}, 
  and with the scenario of transformation of galaxy morphology and 
  luminosity class through galaxy-galaxy interactions/merging \citep{park08}.

On the other hand, we found no dependence of $f_{\rm (U)LIRGs}$ on
  the background density (see Figs. \ref{fig-irsig1d}-\ref{fig-irsig1dl2}).
When we use the distance to the nearest galaxy cluster instead of using $\Sigma_5$,
  the results are similar: dependence of $f_{\rm (U)LIRGs}$
  on the clustercentric distance is weak except for highest-density regions
  such as the center of galaxy clusters.
If the SFA of IRGs is increased by interactions in general, one would expect
  to find an increase of $f_{\rm (U)LIRGs}$ with the background density.
The fact that such trend is not seen is explained by the finding of
  a selective effect of the neighboring galaxy when it is closer than the virial radius
  of the neighbor, i.e. only late types produce an increase of $f_{\rm (U)LIRGs}$.
This can be understood by the need to provide cold gas together with the dynamical interaction.
These results are compatible with the result from the study of Hickson compact groups of galaxies
  that specific SFRs of galaxies in dynamically `young (dominated by late-type galaxies)' groups are, on average, 
  larger than those in dynamically `old (dominated by early-type galaxies)' groups \citep{bit10}.

\begin{figure*}
\center
\includegraphics[scale=0.8]{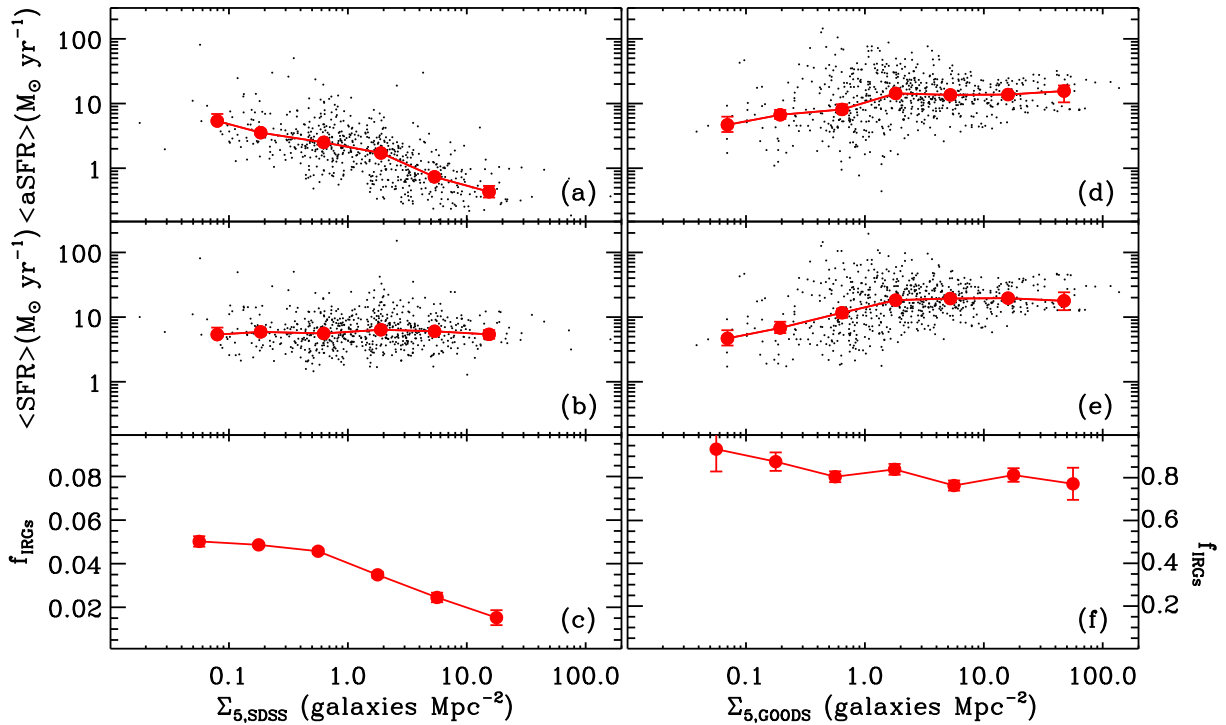}
\caption{(a) Average SFR computed using all galaxies in L1, $<\rm aSFR>$,
  (b) average SFR computed using IRGs, $<\rm SFR>$, and
  (c) the fraction of IRGs among all galaxies as a function of $\Sigma_5$.
(d-f) Same as (a-c), but for the high-redshift IRGs ($0.8<z<1.2$) studied in \citet{elb07}.
Median value and its error at each bin is shown by filled circle and its error-bar.
}\label{fig-sfrenv2}
\end{figure*}

As we mentioned in \S\ref{parameter},
  the sample of (U)LIRGs used in this study
  is dominated by LIRGs. 
Thus these findings demonstrate that not only ULIRGs but also LIRGs
  are strongly affected by interactions/merging.
In order to investigate what makes LIRGs and ULIRGs different in detail,
we plot the ratio of the stellar mass between IRGs ($M_{\rm IRG}$) and 
  their nearest neighbor galaxy ($M_{\rm n}$)
  (i.e. max[$M_{\rm IRG}$,$M_{\rm n}$]/min[$M_{\rm IRG}$,$M_{\rm n}$])
  as a function of $L_{\rm IR}$ in Fig. \ref{fig-ratio}a.
We use all galaxies in L1 or L2 to find the nearest neighbor galaxy of IRGs
  without using the magnitude condition ($M_r=M_{r,\rm target}+0.5$) introduced in \S \ref{env}
  in order to have no constraints on the mass range of neighbor galaxies.
We show only IRGs with $M_r\leq-21$
  in order to reduce the effect of absolute magnitude limit in L1 ($M_r$=$-19.91$)
  on the lower limit of neighbor's mass.
In addition, we plot only IRGs interacting with a late-type nearest neighbor 
  (i.e. IRGs with a late-type neighbor at $R_{\rm n}\leq r_{\rm vir,nei}$). 
It shows that the mass ratio between an IRG and its interacting pair
  appears to decrease as $L_{\rm IR}$ increases.
The fraction of minor merger system (i.e. mass ratio$\geq4$) is found to be
  36, 34, and 14\% for LLIRGs plus MLIRGs, LIRGs and ULIRGs, respectively.
Due to the small number of ULIRGs in our sample,
  the fraction of minor merger system for ULIRGs is computed 
  by including galaxies with $L_{\rm IR}>9\times10^{11} L_\odot$.
We check the statistical significance of the difference in the fraction of minor merger system
  depending on $L_{\rm IR}$ by Monte Carlo simulations.
We construct 1000 trial data sets by randomly selecting the same number of galaxies at each $L_{\rm IR}$ bin
  from the real data, and compute the fraction of simulated data sets 
  in which the median value of mass ratio is larger than 4 (i.e. minor merger system).
We find 17, 18, and 2\% for LLIRGs plus MLIRGs, LIRGs and ULIRGs, respectively.
These results indicate that non-negligible fraction of LIRGs is related to 
  the minor merger/interactions
  between two galaxies with a large mass difference, while
  few ULIRGs are minor merger systems
  (\citealt{ish04,wang06,das06}; see also \citealt{shi06,shi09,elb07,lotz08,ide09}).
 
Using the same sample as above, we show projected distances to 
  the nearest neighbor galaxy of IRGs
  as a function of $L_{\rm IR}$ in Fig. \ref{fig-ratio}b.
It is seen that distances to the nearest neighbor galaxies are
  smaller for ULIRGs than those for LIRGs.
If IRGs with $R_{\rm n}\leq 0.1r_{\rm vir,nei}$ can be regarded as 
  advanced merger systems,
  the fraction of advanced merger systems for LLIRGs plus MLIRGs, LIRGs and ULIRGs
  is 6, 21, and 50\%, respectively.
We check the statistical significance of the difference in median values of $R_{\rm n}/r_{\rm vir,nei}$ 
  depending on $L_{\rm IR}$ based on the Monte Carlo simulations with 1000 trial data sets made
  by randomly selecting the same number of galaxies at each $L_{\rm IR}$ bin
  from the real data.
We compute the fraction of simulated data sets 
  in which the median value of $R_{\rm n}/r_{\rm vir,nei}$ 
  is less than 0.1 (i.e. advanced merger), and
  find 0, 3, and 49\% for LLIRGs plus MLIRGs, LIRGs and ULIRGs, respectively.
These results can suggest that ULIRGs are more advanced merger systems 
  (or strong interaction) than LIRGs.

In summary, the SFA of both LIRGs and ULIRGs is triggered by the 
  interaction/merging with late-type (gas-rich) neighbor galaxies,
  but they seem to differ in the sense that 
  1) stellar masses of ULIRGs are larger than those of LIRGs 
  (see Figs. \ref{fig-sfrir}c and \ref{fig-irneib2da}),
  2) few ULIRGs are minor merger systems, but
    a non-negligible fraction of LIRGs ($\sim$34\%) is a minor merger system, and
  3) ULIRGs are more advanced merger systems than LIRGs.

\subsection{Evolution of SFR-environment Relation}

To investigate how the effects of the environment on the SFA of galaxies 
  evolve with cosmic time, we plot, in Figs. \ref{fig-sfrenv1} and \ref{fig-sfrenv2}, 
  several SFA parameters of IRGs
  as a function of the background density ($\Sigma_5$) 
  in comparison with those at high redshifts studied in \citet{elb07}.

In Fig. \ref{fig-sfrenv1},
  SFRs, stellar masses, and specific SFRs for local IRGs (a-c) 
  do not show any significant change along with the background density,
  while SFRs and stellar masses for high redshift IRGs (d-e) increase
  as the background density increases.

In Fig. \ref{fig-sfrenv2},
  we show the spatially averaged SFRs for IRGs  
  as a function of the background density.
The average SFR is computed at the position of IRG
  using the neighbor galaxies that have velocities relative to the IRG less than 
  1500 km s$^{-1}$ and the projected distance to the IRG less than 1 Mpc.
$<\rm aSFR>$ in (a and d) is the average SFR computed using all neighbor galaxies
  regardless of IR detection, while $<\rm SFR>$ in (b and e) is the one computed 
  using the neighbor IRGs. 
For the GOODS galaxies studied in \citet{elb07},
  we first compute the background density ($\Sigma_{\rm 5,GOODS}$)
  by identifying 5th-nearest neighbor of each IRG
  in the spectroscopic sample of galaxies with $0.8<z<1.2$ and $z_{AB}\leq23.5$
  (see \citealt{elb07} for details of the data).
Secondly, we compute the average SFRs adopting the same method as above.

As seen in (a) and (d), the average SFR ($<\rm aSFR>$) for local SDSS IRGs
  decreases as the background density increases, 
  while that for high redshift IRGs increases 
  as the background density increases.
This is the reversal of the SF-density relation in the distant universe
  found by \citet{elb07}.
The different dependence of the average SFR on the background density
  might be caused by the different fraction of star-forming galaxies
  depending on the background density (see c and f):
  since the fraction of star-forming galaxies in the local universe
  is very small in high-density regions, the average SFRs in these regions
  are also small due to the large contribution of non-SF galaxies.

If we see the behavior of the average SFR ($<\rm SFR>$) computed 
  using the neighbor IRGs, $<\rm SFR>$ for local IRGs in (b) 
  appears not to depend on the background density, 
  while $<\rm SFR>$ for high redshift galaxies in (e)
  still shows an increase along with the background density.
Since the average SFR ($<\rm SFR>$) computed using the neighbor IRGs
  is not affected by the non-SF galaxies, it may provide important hints
  for what directly affects the SFA.
Since the SFR of a galaxy shows a strong correlation with the galaxy mass
  as seen in \S \ref{results} (see also \citealt{bri04,elb07,noe07mass,noe07sf}),
  it is instructive to see the mass dependence of IRGs on the background density.
It is seen in Fig. \ref{fig-sfrenv1}(b) and (e) that 
  the median stellar masses for local IRGs do not show 
  a significant change with the background density, 
  while those for high redshift galaxies 
  increase as the background density increases, which are responsible for 
  the difference of the average SFRs ($<\rm SFR>$) between Fig. \ref{fig-sfrenv2}(b) and (e).

Finally, the key in the SF-density relation seems to be how the fraction
  of massive, star-forming galaxies changes with the background density.
Then the reversal of SF-density relation can be understood by the idea
  that massive, star-forming galaxies are strongly clustered at high redshifts
  (e.g., \citealt{far06,gil07})
  as forming a larger structure such as a galaxy group/cluster,
  and evolve (or consume their gas) through galaxy-galaxy interactions/merging 
  faster than those in the field.
Thus, gas-exhausted, non-SF galaxies are preferentially 
  found in high-density regions of the local universe.

This environmental dependence of evolution of the SFA of galaxies 
  is consistent with the results of recent studies of other galaxy properties 
  such as galaxy morphology (e.g., \citealt{cap07,hp09}).
For example, \citet{hp09} found that
  the early-type fraction evolves much faster in high-density regions
  than in low-density regions, and that the morphology-density relation
  becomes weaker at $z\approx1$ (see their Fig. 8).
It may be because the rate of galaxy-galaxy interactions is higher 
  in high-density regions,
  and a series of interactions/merging over the course of galaxy life
  consume most of its cold gas and/or transform late types into early types.

\section{Conclusions}\label{con}

By cross-correlating the spectroscopic sample of galaxies in the SDSS DR7
  and the IRAS FSC92,
  we obtained a sample of nearby LIRGs and studied their environmental dependence. 
We examined the effects of the large-scale background density,
  galaxy clusters, and 
  the nearest neighbor galaxy on the properties of IRGs.
Our main results are as follows.

\begin{enumerate}
\item We find no dependence of the fraction of LIRGs plus ULIRGs 
  among IRGs ($f_{\rm (U)LIRGs}$) on the background density ($\Sigma_5$),
  but find a strong dependence on stellar mass. 

\item When we use the distance to the nearest galaxy cluster 
  instead of using $\Sigma_5$, the results are similar to
  those based on $\Sigma_5$:
  the dependence of $f_{\rm (U)LIRGs}$ on the distance to the nearest galaxy cluster
  is weak except for the highest-density region such as the center of galaxy clusters.
  
\item We find that $f_{\rm (U)LIRGs}$ and $L_{\rm IR}$ of IRGs  
  strongly depend on the morphology of and the distance to the nearest neighbor galaxy.
When an IRG has a close neighbor galaxy,
  the probability for it to be a (U)LIRG ($f_{\rm (U)LIRGs}$) and its $L_{\rm IR}$
  increase as it approaches a late-type neighbor, but
  decrease as it approaches an early-type neighbor.
The bifurcations of $L_{\rm IR}$ and $f_{\rm (U)LIRGs}$ depending on the neighbor's morphology
    are seen at $R_n\approx 0.5 r_{\rm vir,nei}$.
    
\item The median $L_{\rm IR}$ values of IRGs and $f_{\rm (U)LIRGs}$ 
    at the largest neighbor separations seem to be
    larger than those at the intermediate neighbor separation.
This supports the idea that (U)LIRGs are powered by the interaction/merging with 
  late-type (gas-rich) nearest neighbor galaxies as expected for non-resolved
  advanced mergers.

\item LIRGs and ULIRGs are found to differ in the sense that 
  1) stellar masses of ULIRGs are larger than those of LIRGs,
  2) few ULIRGs are minor merger systems, but
    a non-negligible fraction of LIRGs ($\sim$34\%) is a minor merger system, and
  3) ULIRGs are more advanced merger systems than LIRGs.

\item The environmental dependence of (U)LIRGs and the evolution of the SFR-environment
  relation from high redshifts to low redshifts seem to be consistent with the idea that 
  galaxy-galaxy interactions/merging play a critical role in triggering the SFA of (U)LIRGs.
Massive, star-forming galaxies at high redshifts are strongly clustered,
  and evolve (or consume their gas) through galaxy-galaxy interactions/merging 
  faster than those in the field.
Thus, gas-exhausted, non-SF galaxies are preferentially 
  found in high-density regions of the local universe.

\end{enumerate}

\begin{acknowledgements}
We would like to thank the anonymous referee for constructive comments that 
  helped us to improve the manuscript.
We also thank C. Feruglio for helpful discussions.
CBP acknowledges the support of the National Research Foundation of Korea (NRF) grant
  funded by the Korea government (MEST) (No. 2009-0062868).
MGL is supported in part by a grant (R01-2007-000-20336-0) 
  from the Basic Research Program of the Korea Science and Engineering Foundation.
Funding for the SDSS and SDSS-II has been provided by the Alfred P. Sloan 
Foundation, the Participating Institutions, the National Science 
Foundation, the U.S. Department of Energy, the National Aeronautics and 
Space Administration, the Japanese Monbukagakusho, the Max Planck 
Society, and the Higher Education Funding Council for England. 
The SDSS Web Site is http://www.sdss.org/.
The SDSS is managed by the Astrophysical Research Consortium for the 
Participating Institutions. The Participating Institutions are the 
American Museum of Natural History, Astrophysical Institute Potsdam, 
University of Basel, Cambridge University, Case Western Reserve University, 
University of Chicago, Drexel University, Fermilab, the Institute for 
Advanced Study, the Japan Participation Group, Johns Hopkins University, 
the Joint Institute for Nuclear Astrophysics, the Kavli Institute for 
Particle Astrophysics and Cosmology, the Korean Scientist Group, the 
Chinese Academy of Sciences (LAMOST), Los Alamos National Laboratory, 
the Max-Planck-Institute for Astronomy (MPIA), the Max-Planck-Institute 
for Astrophysics (MPA), New Mexico State University, Ohio State University, 
University of Pittsburgh, University of Portsmouth, Princeton University,
the United States Naval Observatory, and the University of Washington. 
This research has made use of the NASA/IPAC Extragalactic Database (NED) 
which is operated by the Jet Propulsion Laboratory, California Institute of Technology, 
under contract with the National Aeronautics and Space Administration.
\end{acknowledgements}

\bibliographystyle{aa} 
\bibliography{ref_hshwang} 
\end{document}